\documentclass[aps,prl,reprint,superscriptaddress,showpacs]{revtex4-1}
\usepackage{mathtools}
\usepackage{amssymb}
\usepackage{amsthm}
\usepackage{amsfonts}
\usepackage{graphicx}
\usepackage[usenames,dvipsnames]{xcolor}
\usepackage{tabularx}
\usepackage[normalem]{ulem}
\usepackage{microtype}
\usepackage{enumerate}
\usepackage{hyperref}
\usepackage{bookmark}
\usepackage{tikz}
\usepackage{tikz}
\usepackage{pgfplots}
\usetikzlibrary{plotmarks}

\hypersetup{
 pdftitle={The six-vertex model and Schramm-Loewner evolution},
 pdfauthor={Richard Kenyon, Jason Miller, Scott Sheffield, and David B. Wilson},
colorlinks=true,
linkcolor=NavyBlue,
urlcolor=Blue,
citecolor=PineGreen,
}

\newcommand{\old}[1]{}
\newcommand{\SLE}{\operatorname{SLE}}
\newcommand{\CLE}{\operatorname{CLE}}
\newcommand{\Z}{{\mathbb Z}}
\newcommand{\G}{{\mathcal G}}

\newcommand{\rO}{\mathrm{O}}

\begin{document}

\title{The six-vertex model and Schramm--Loewner evolution}

\author{Richard Kenyon}
\affiliation{Brown University}
\author{Jason Miller}
\affiliation{Cambridge University}
\author{Scott Sheffield}
\affiliation{Massachusetts Institute of Technology}
\author{David B.\! Wilson}
\affiliation{Microsoft Research}

\begin{abstract}
Square ice is a statistical mechanics model for two-dimensional ice,
widely believed to have a conformally invariant scaling limit.
We associate a Peano (space filling) curve to a square ice configuration,
and more generally to a so-called $6$-vertex model configuration, and 
argue that its scaling limit is a space-filling version of the random
fractal curve $\SLE_{\kappa}$, Schramm--Loewner evolution with
parameter $\kappa$, where $4<\kappa\leq 12+8\sqrt{2}$.
For square ice, $\kappa=12$.  At the ``free-fermion point'' of the
6-vertex model, $\kappa=8+4\sqrt{3}$.  These unusual values lie
outside the classical interval $2\le \kappa\le 8$.
\end{abstract}

\pacs{64.60.De, 64.60.al}
\maketitle

\begin{figure}[t!]
\vspace{-9.5pt}
\hspace{0pt}\hfill
\clap{\includegraphics[width=.24\textwidth]{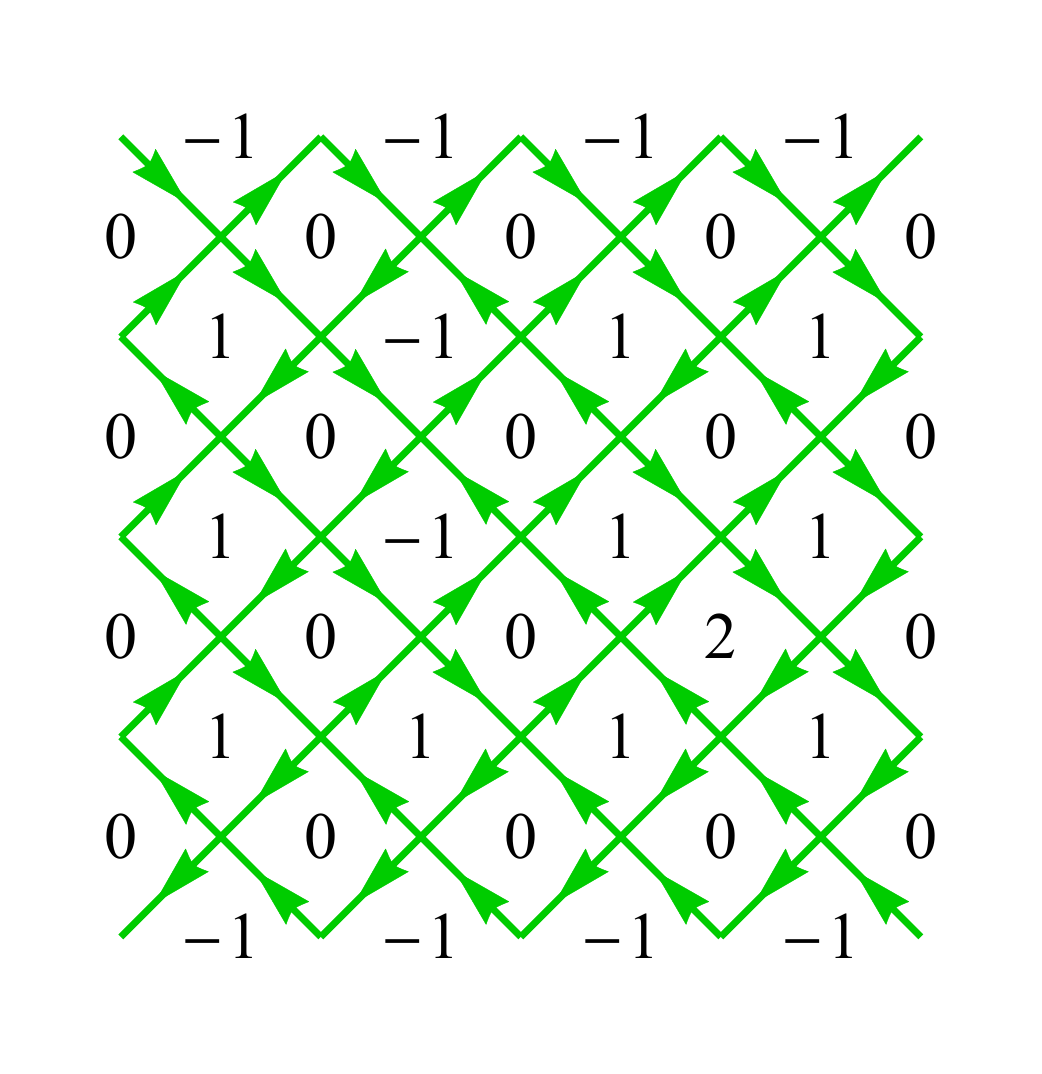}}\clap{(a) 6-vertex configuration}\hfill\hfill
\clap{\includegraphics[width=.24\textwidth]{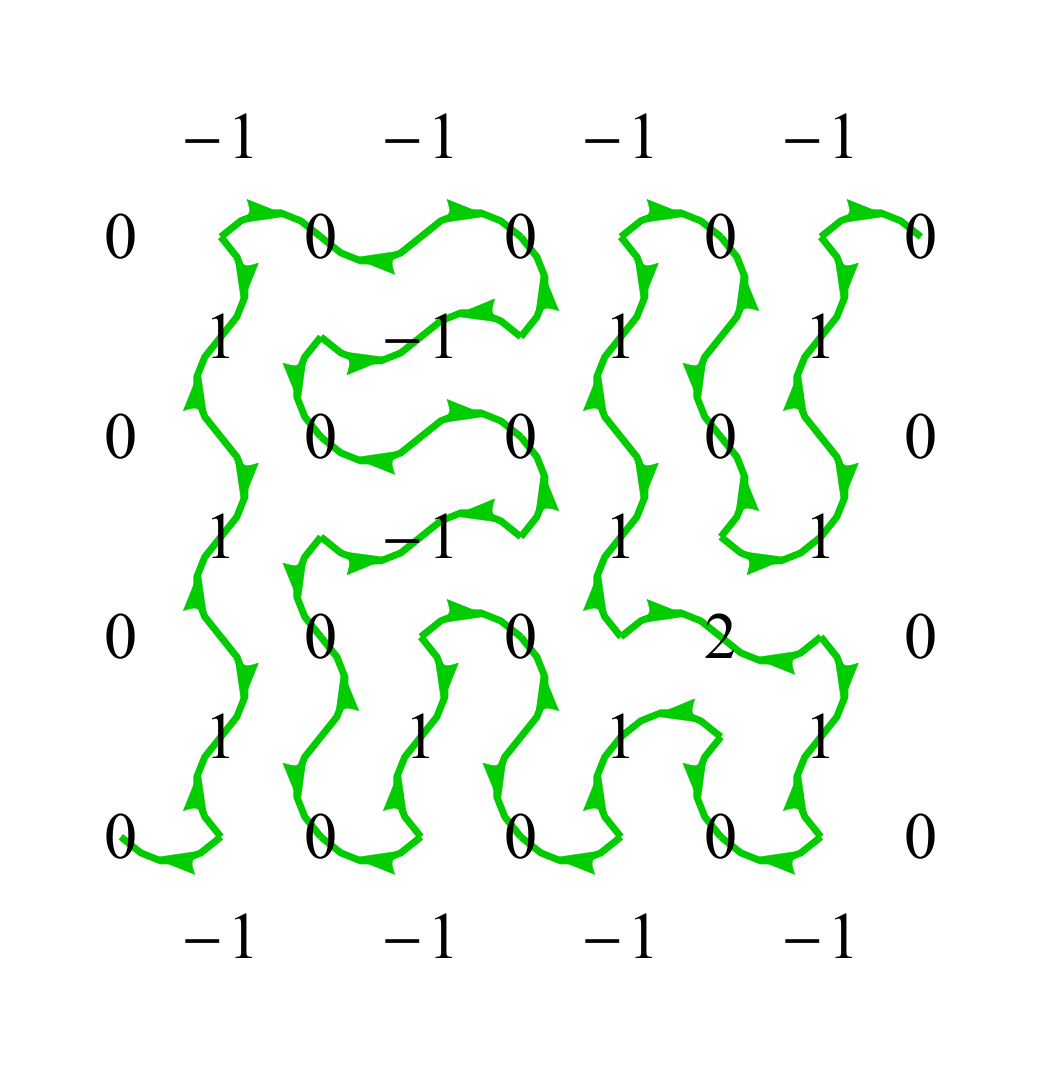}}\clap{(b) Peano curve}
\hfill\hspace{0pt}
\caption{6-vertex configuration with height function and its Peano curve.
The heights, when scaled by $\pi/2$, give the winding of the Peano curve up to an additive constant.
}
\label{fig:ice-tours}
\end{figure}

\bookmark[dest=introduction]{Six-vertex model introduction}
\hypertarget{introduction}{}
Square ice was introduced by Pauling \cite{pauling:ice} as a model of hydrogen bonding
in ice crystals in two dimensions \cite{baxter}.
A square-ice configuration is an orientation of each edge of the square lattice, 
subject to the constraint that each vertex has two incoming and two outgoing edges
(see the diagram below and Fig.~\ref{fig:ice-tours}). 
Recently actual square ice crystals were produced between sheets of graphene \cite{square-ice}.

The classical $6$-vertex model from statistical mechanics generalizes square ice by adding energies to each of the $6$ types of local configuration at a vertex:
\begin{figure}[h]
\vspace{-6pt}
\includegraphics[width=.5\textwidth]{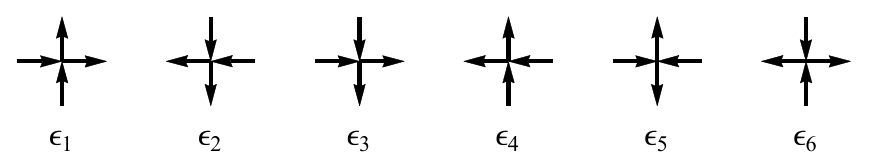}
\vspace{-14pt}
\end{figure}

\noindent
Square ice is the uniform measure on 6-vertex configurations.
The 6-vertex model partition function was famously
solved by Lieb in 1967 \cite{lieb:ice}. A number of beautiful combinatorial identities arising in this model 
have intrigued mathematicians
and physicists for many years \cite{razumov-stroganov:ASM,cantini-sportiello}. In particular
it is widely believed that the $6$-vertex model has conformally invariant
scaling limits, however a mathematical proof of this fact is lacking.

We show here how to associate a discrete Peano (space filling) curve to
configurations of the square ice model with appropriate boundary conditions
(Fig.~\ref{fig:ice-tours}).
We present evidence that the scaling limit of this curve is a random
fractal curve called a Schramm--Loewner evolution (SLE).

For each $\kappa \leq 0$, an $\SLE_\kappa$ in the upper half plane is a random non-self-crossing random curve that extends from the origin to $\infty$, with the parameter $\kappa$ indicating how ``windy'' the path is.  In recent decades, SLE has been thoroughly studied and celebrated within both physics and mathematics, and has led to many new results about two-dimensional statistical physics and the Liouville theory of quantum gravity --- some of which go far beyond the results previously established using conformal field theory and other techniques.

The precise definition of SLE is interesting and indirect.
Fix $\kappa > 0$, let $B(t)$ be a one-dimensional Brownian motion, and for each $z$ in the complex upper half plane $\mathbb H$, let $g_t(z)$ solve the ODE
\[\frac{\partial g_t(z)}{\partial t}= \frac{2}{g_t(z)-\sqrt{\kappa}\, B(t)}\quad\quad\quad\quad g_0(z) = z\,,\]
which is defined until $T_z = \inf \{t : g_t(z)-W_t = 0 \}$. Then $\SLE_\kappa$ is the curve $\eta: \mathbb R_+ \to \mathbb H$ defined so that $\{z : T_z \leq t\}$ is the set of points hit or cut off from $\infty$ by $\eta([0,t])$.

For $\kappa\leq 4$, $\SLE_\kappa$ is a simple curve;
for $4<\kappa<8$, the curve hits itself without crossing itself,
forming bubbles;
for $\kappa\geq 8$, the curve is space-filling \cite{rohde-schramm}.
For $4<\kappa<8$, there is also a space-filling version of $\SLE_\kappa$
in which the bubbles get filled in recursively as they are made
\cite{miller-sheffield:ig4}.

The $\SLE_\kappa$ curves are either known or
believed to characterize the scaling limits of various
two-dimensional critical statistical physics models: 
dilute polymers ($\kappa=8/3$) \cite{LSW:SAW}, dense polymers
($\kappa=8$) \cite{LSW:tree}, loop-erased random walk ($\kappa=2$)
\cite{LSW:tree}, percolation interfaces ($\kappa=6$)
\cite{smirnov:percolation}, Ising model spin clusters ($\kappa=3$)
\cite{smirnov:ising,CDCHKS:ising}, dimer systems ($\kappa=4$),
contours of the Gaussian free field ($\kappa=4$)
\cite{schramm-sheffield:discrete-GFF,schramm-sheffield:continuum-GFF},
the Ashkin--Teller model ($\kappa=4$), the Fortuin--Kasteleyn random
cluster model ($2\leq\kappa\leq 8$),
active spanning trees ($4<\kappa\leq 12$) \cite{kassel-wilson:active},
and others.  The dimension $D_{\!f}$ of the
fractal increases with the parameter $\kappa$ according to the
formula $D_{\!f}=\min(2,1+\kappa/8)$ \cite{rohde-schramm,Beffara}.
See \cite{rohde-schramm,Cardy,Schramm-ICM} for further background.

SLE is connected with conformal field theory (CFT) \cite{Cardy}, where the central charge $c$ is related to $\kappa$ by
\begin{equation} \label{eqn:c-kappa}
c = (8 - 3\kappa) (\kappa - 6)/(2 \kappa)\,.
\end{equation}
In CFT usually $c\geq-2$, which corresponds to $\kappa \in [2,8]$, the values relevant to {\em conformal loop ensembles} \cite{sheffield:cle-trees}. 
Before this work and \cite{kassel-wilson:active} it was widely assumed that only $\kappa \in [2,8]$ would appear in natural discrete models \cite{Cardy}.

For the 6-vertex model Peano curve defined here,
$\kappa$ depends on the vertex energies and spans the range
$(4, 12+8\sqrt{2}]$, which in particular includes values outside of $[2,8]$.
For square ice, $\kappa=12$, which corresponds to $c=-7$.
The square ice Peano curve
joins a tiny pantheon of models (including the uniform spanning tree and the Ising model) that have \textit{independently\/} solvable random lattice analogs; these analogs are described in \cite{KMSW1}, along with connections to Liouville quantum gravity and string theory.

\medskip
\bookmark[dest=height-tour]{Heights and Peano curve}
\hypertarget{height-tour}{\noindent\textit{\textbf{6-vertex model height function and Peano curve.}}}
Six-vertex configurations have a height function which plays an
important role in their analysis \cite{vanbeijeren:roughening}.  The
heights are defined on the faces; around even-parity vertices, the
heights increase by $1$ in the counterclockwise direction across
outgoing edges, and decrease by $1$ in the counterclockwise direction
across incoming edges (see Fig.~\ref{fig:ice-tours}).

\begin{figure}[b]
\vspace{-8pt}
\includegraphics[width=2.5in]{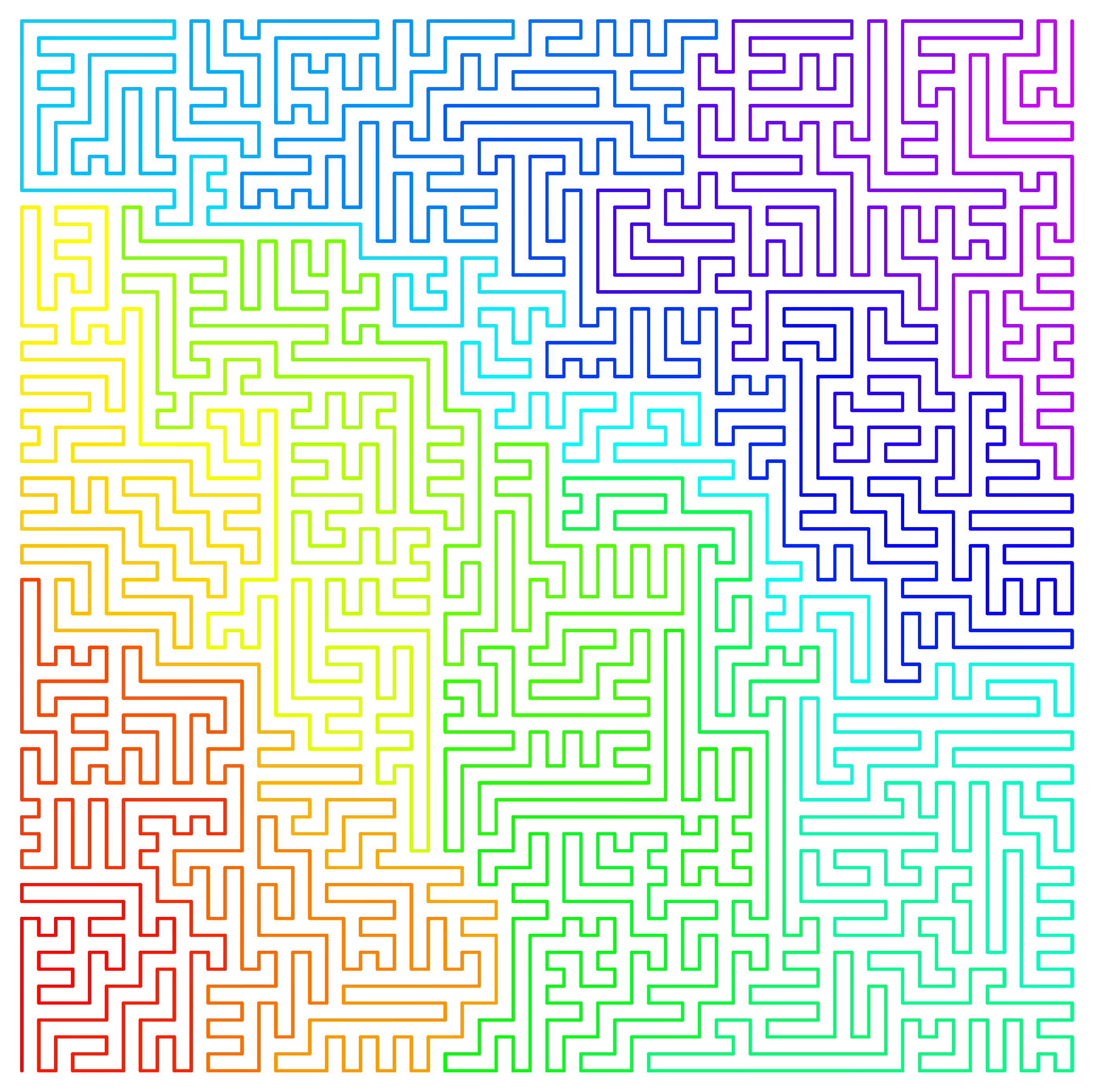}
\caption{\label{fig:ice-tour}The Peano curve, colored according to the time parameter, for the square ice model ($C=1, \kappa'=12$).}
\end{figure}

We produce a Peano curve (Figs.~\ref{fig:ice-tours} and~\ref{fig:ice-tour})
from a 6-vertex configuration as follows:
From the even index vertices, we bend the outgoing arrows $45^\circ$
left so that they terminate at the face centers, and from the odd index
vertices, we bend the outgoing arrows $45^\circ$ right.  Note that each arrow
gets bent into the same face regardless of which way it is oriented,
and each face receives two arrows from opposite sides.
Because each face and each vertex now has degree two, the curved arrows
form a collection of loops and chains which terminate at the boundary.
Observe that the six-vertex heights, when scaled by $\pi/2$, give the
winding angle of the green curve measured in radians.  Because the
height function is single-valued, the green curve cannot close up on itself
to form loops.  The boundary conditions were chosen so that there is only one
chain, so it must form a single space-filling curve.

\medskip
\bookmark[dest=O(n)]{6-vertex and O(n) loop models}
\hypertarget{O(n)}{\noindent\textit{\textbf{6-vertex and \texorpdfstring{$\mathbf{O}(n)$}{O(n)} loop models.}}}
The six-vertex model can also specialize to the $\rO(n)$ loop model.  To obtain the $\rO(n)$ model, we
set (with $\omega_i=e^{-\epsilon_i}$)
$\omega_1=\omega_2=\omega_3=\omega_4=1$ and $\omega_5=\omega_6=C$.  The parameter $\Delta$ is defined by
\begin{equation} \label{eqn:Delta}
 \Delta = \frac{\omega_1\omega_2+\omega_3\omega_4-\omega_5\omega_6}{2\sqrt{\omega_1\omega_2\omega_3\omega_4}}=\frac{2-C^2}{2}.
\end{equation}

There is a weight-preserving mapping between six-vertex configurations and $\rO(n)$ model
loop configurations, so that the partition functions are equal \cite{BKW}:
One splits each vertex in half (maintaining planarity) so that each half has one out-going and
one in-coming edge.  For any vertex with adjacent out-going arrows, there is one way to do this split,
but for $C$-type vertices, there are two ways to split it.
A split vertex is given a weight of $r$ if the arrows turn right, and weight
$1/r$ if the arrows turn left.  For the non-$C$-type vertices, the total weight is $r\times r^{-1} = 1$.  For the
$C$-type vertices, the total weight is $r^2+ r^{-2} = C$.  Each loop has weight $r^4 + r^{-4} = n$.
Thus
\begin{equation} n = C^2 - 2 \label{eqn:n-C}\end{equation}
and hence $n=-2\Delta$.

The $\rO(n)$ model loops are widely believed to be described by the conformal loop ensemble $\CLE_{\kappa^\circ}$ (the loop version of SLE), where
\begin{equation}
n = -2\cos(4\pi/\kappa^\circ) \label{eqn:n-kappa}
\end{equation}
\cite{sheffield:cle-trees}.  (Here ${}^\circ$ is a mnemonic for $\rO(n)$.)
The $\SLE$-parameter for the Peano curve coming from the associated 6-vertex model we call $\kappa'$.
Interestingly, $\kappa'\neq\kappa^\circ$.

\newcommand{\sinklab}{\llap{\raisebox{23pt}{\footnotesize\clap{\colorbox{white}{\ }}\clap{\color{white}\textbf{sink}}\clap{sink}\hspace{100pt}}}}
\newcommand{\sourcelab}{\llap{\raisebox{88pt}{\footnotesize\clap{\color{white}\textbf{source}}\clap{source}\hspace{10pt}}}}
\begin{figure*}[t!]
\vspace{-10pt}
\hspace{0pt}%
\rlap{\includegraphics[width=1.6in]{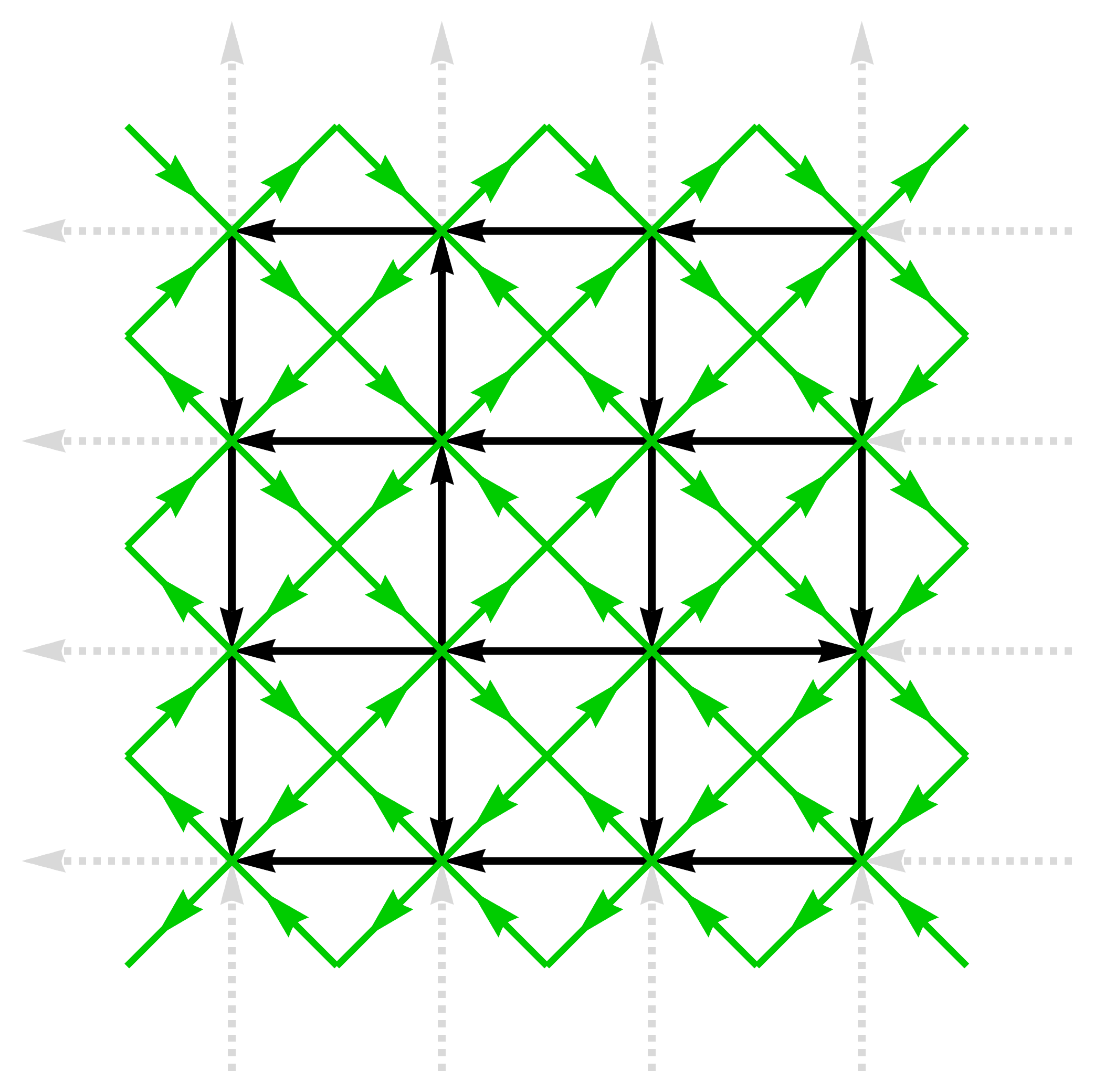}\sinklab\sourcelab}\hspace{.8in}\clap{(a)}\hspace{.8in}\hfill
\rlap{\includegraphics[width=1.6in]{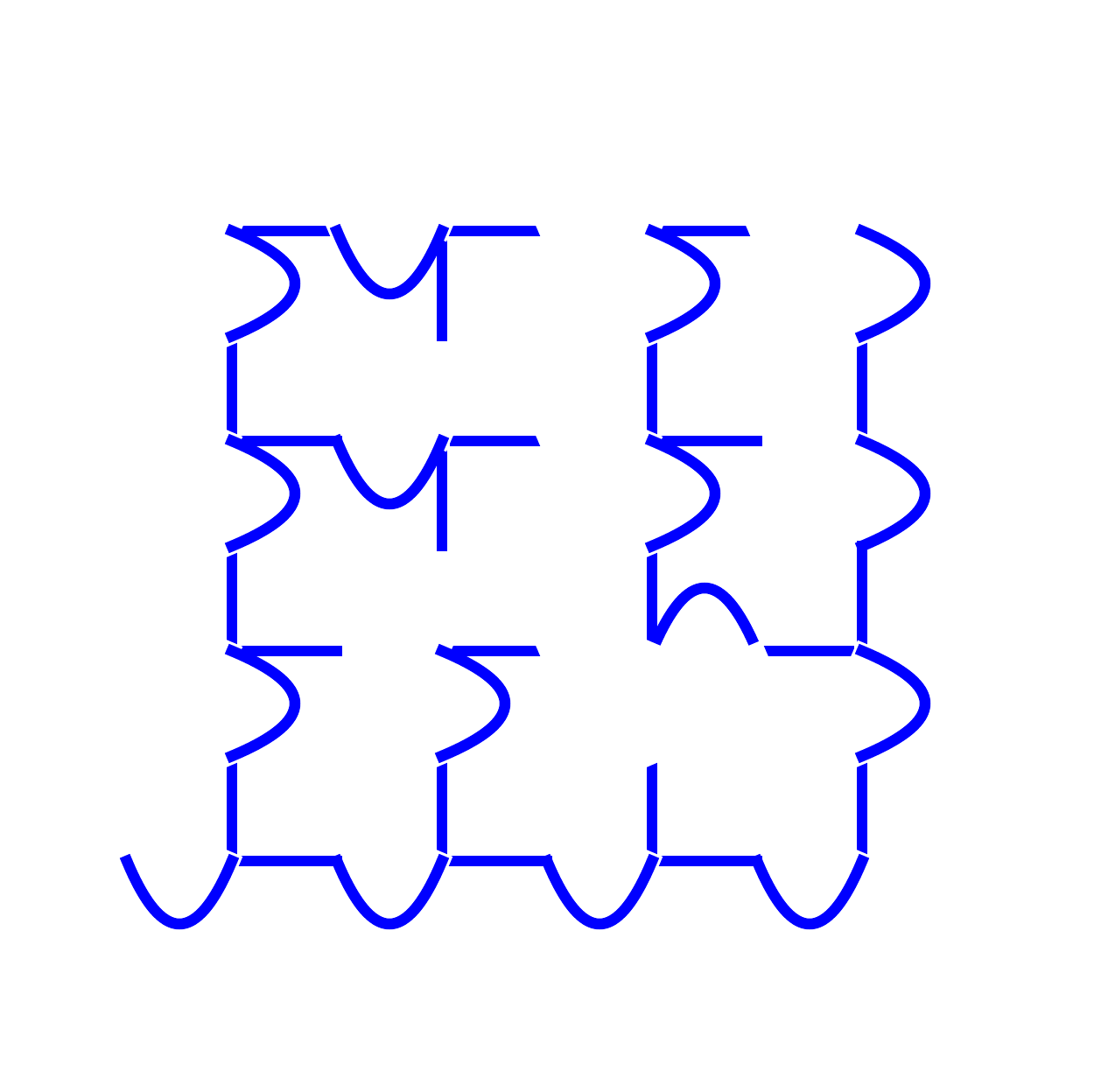}\sinklab}\hspace{.8in}\clap{(b)}\hspace{.8in}\hfill
\rlap{\includegraphics[width=1.6in]{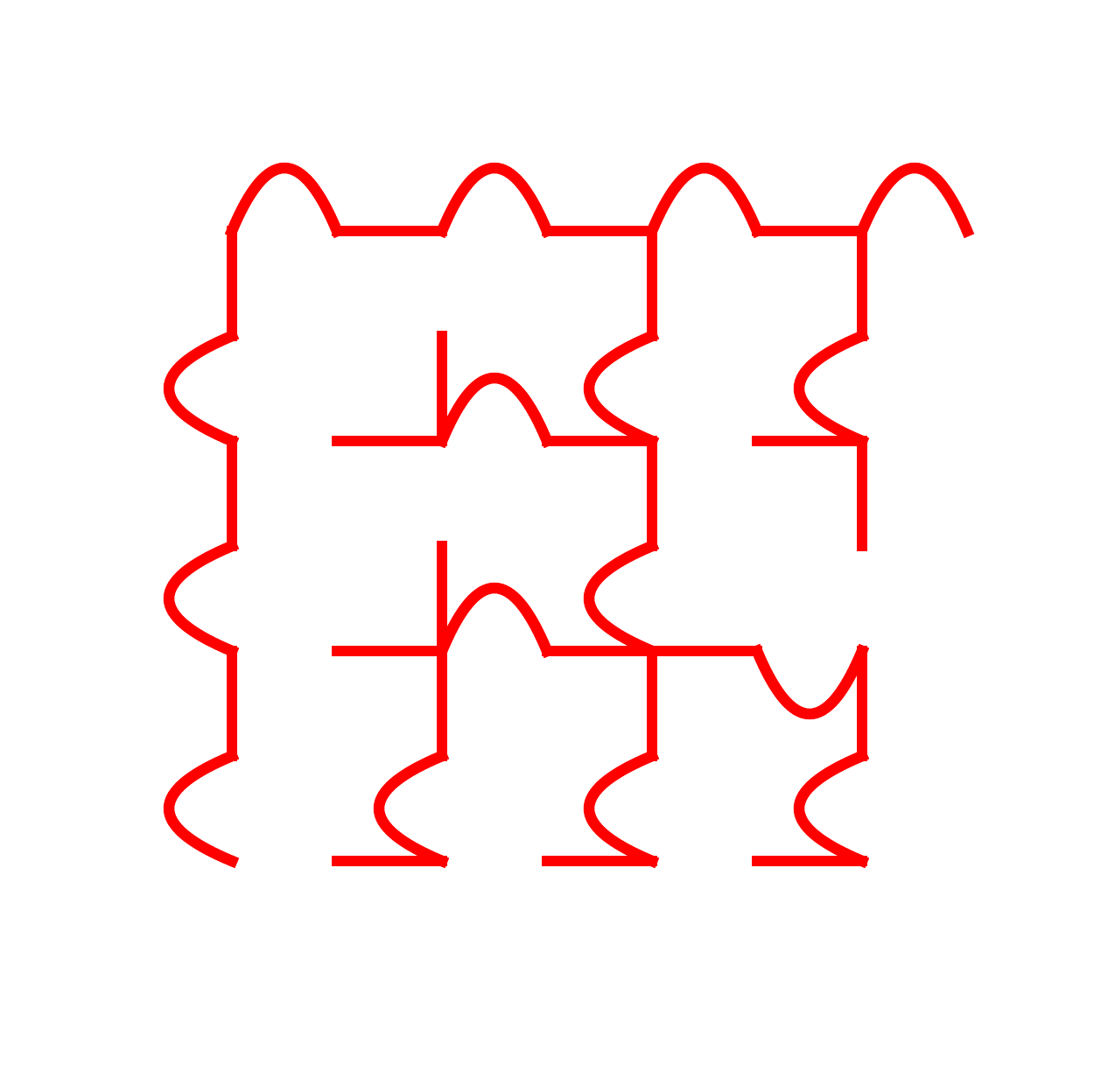}\sourcelab}\hspace{.8in}\clap{(c)}\hspace{.8in}\hfill
\rlap{\includegraphics[width=1.6in]{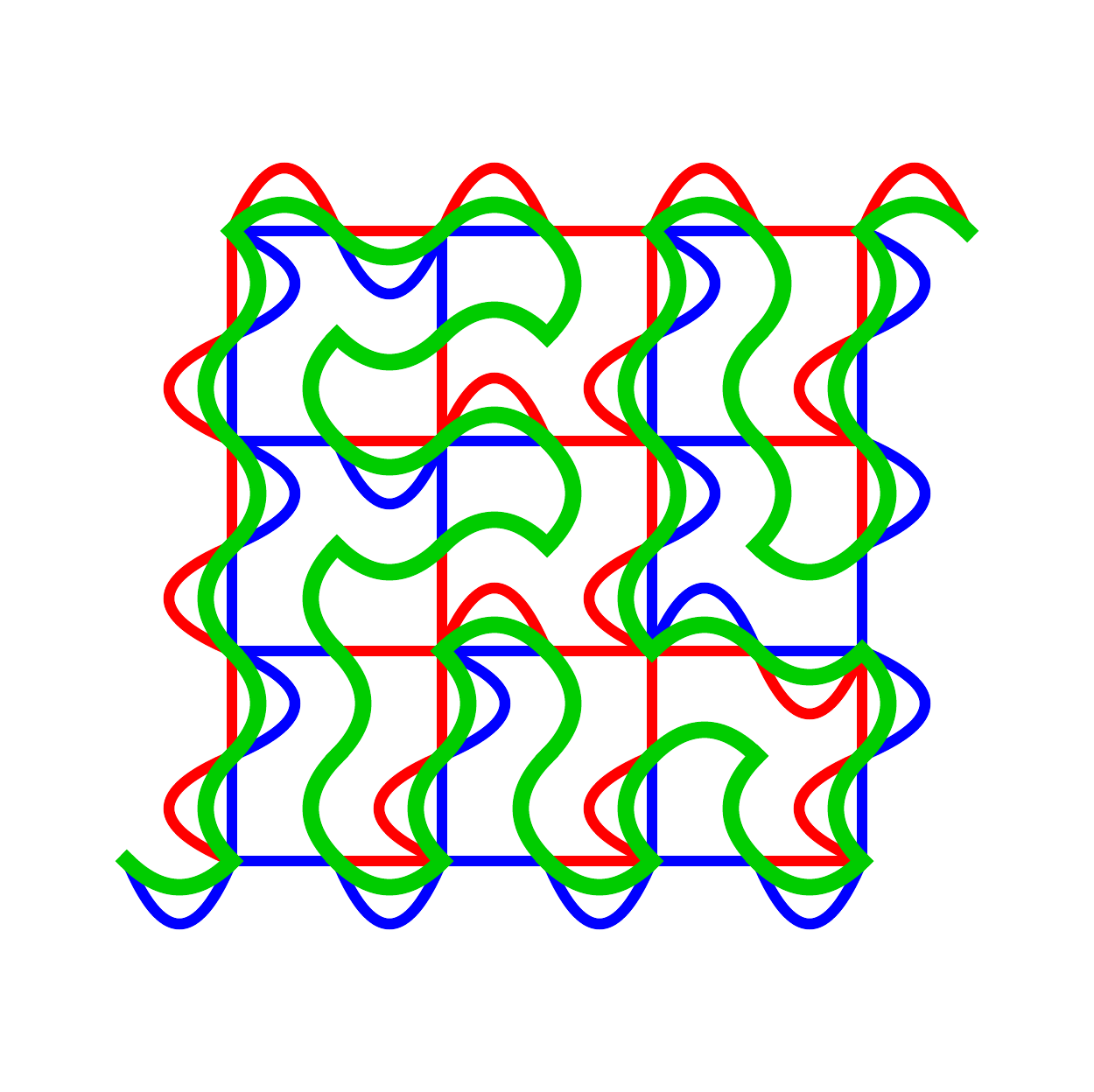}}\hspace{.8in}\clap{(d)}\hspace{.8in}
\caption{Shown here are (a) square ice (green) with bipolar orientation (black), (b) SE-tree (blue) rooted at sink, (c) NW-tree (red) rooted at source, and (d) SE-tree and NW-tree (which do not cross each other) with Peano curve separating them (green).  In the SE-tree and NW-tree, each edge leads to a vertex, drawn with a straight segment, and each vertex leads to an edge, drawn with a curved segment.  The Peano curve separating the SE-tree and NW-tree is the same as the Peano curve defined from the six-vertex heights in Fig.~\ref{fig:ice-tours}.}
\label{bpfig}
\end{figure*}

\medskip
\bookmark[dest=height-variance]{Height function variance}
\hypertarget{height-variance}{\noindent\textit{\textbf{6-vertex height function variance.}}}
The variance in the height function of the six-vertex model was computed
by Nienhuis \cite{nienhuis:6v}:
When the height function $h$ is
measured in radians,
for small $a$,
$\langle\exp(ia(h(x)-h(0)))\rangle = \exp(-a^2/g \log|x|)$, where
$g$ is the Coulomb gas coupling constant.
So the height variance, given by the quadratic term (in $a^2$), is $(1/g)\log|x|$.
From \cite[(3.29)]{nienhuis:6v} we have
\begin{equation} \label{eqn:g-C}
\sin\frac{\pi g}{8} = \frac{C}{2}.
\end{equation}

The theory of imaginary geometry, as developed by Miller and
Sheffield, associates to a Gaussian free field (GFF) a space-filling
SLE \cite{miller-sheffield:ig1,miller-sheffield:ig2,miller-sheffield:ig3,miller-sheffield:ig4}.
Roughly speaking, the GFF height function $h$ is divided by a
parameter $\chi$ to obtain a field of orientations (measured in
radians), and the orientation of the SLE curve is $e^{i h/\chi}$.
Thus the Coloumb gas coupling constant $g$ and the parameter $\chi$ are
(heuristically) related by $g=\chi^2$.

The space-filling SLE parameter $\kappa'$ and $\chi$ are related by
$
\chi = \frac{\sqrt{\kappa'}}{2} - \frac{2}{\sqrt{\kappa'}}
$
\cite{miller-sheffield:ig4}, so
\begin{equation} \label{eqn:g-chi-kappa}
\frac{1}{g} = \frac{1}{\chi^2} = \frac{4\kappa'}{(\kappa'-4)^2}\,.
\end{equation}

If we parametrize $n$ by $n=-2\cos\theta$ with $0\leq\theta\leq\pi$, then
\eqref{eqn:Delta}, \eqref{eqn:n-C}, \eqref{eqn:g-C}, \eqref{eqn:g-chi-kappa}, and \eqref{eqn:c-kappa}
can be expressed as
\begin{align}
n&=-2\cos\theta\notag\\
\Delta&=-\cos\theta\notag\\
C^2 &=2-2\cos\theta\notag\\
\chi^2=g &= 4\,\theta/\pi\notag\\
\kappa'&=4+8\,\theta/\pi+8\sqrt{\theta/\pi+\theta^2/\pi^2}\label{eqn:kappa'}\\
c'&=1-24\,\theta/\pi\,,\notag
\end{align}
where $c'$ is the central charge associated with $\SLE_{\kappa'}$.

The table below gives some special cases.
The limiting case $C\to0$ is included, but with $C=0$ the discrete models do not converge to SLE.
Square ice is the $C=1$ row.
The special value $C=\sqrt{2}$ is the
``free fermion'' point,
where there is a mapping between the 6-vertex model and square-lattice dimers;
in this case $\kappa'=8+4\sqrt{3}$.

\medskip
\noindent\mbox{\begin{ruledtabular}\medskip
\begin{tabular}{ccccccrr}
$\theta$         & $n$         & $\Delta$      & $C$                 & $1/\chi^2$ & $\kappa'$           & $c'$  \\
$\pi$            & $2$         & $-1$          & $2$                 & $1/4$      & $12+8\sqrt{2}$      & $-23$ \\
$\frac{2}{3}\pi$ & $1$         & $-1/2$        & $\sqrt{3}$          & $3/8$      & $28/3+8\sqrt{10}/3$ & $-15$ \\
$\frac{1}{2}\pi$ & $0$         & $0$           & $\sqrt{2}$          & $1/2$      & $8 + 4\sqrt{3}$     & $-11$ \\
$\frac{1}{3}\pi$ & $-1$        & $1/2$         & $1$                 & $3/4$      & $12$                & $-7$  \\
$0$              & $-2$        & $1$           & $0$                 & $\infty$   & $4$                 & $1$   \\
\end{tabular}
\end{ruledtabular}}

\bigskip
\bookmark[dest=bipolar]{Bipolar orientations and space-filling trees}
\hypertarget{bipolar}{\noindent\textit{\textbf{Bipolar orientations and space-filling trees.}}}
There is a useful, and related, bijection between six-vertex configurations and \emph{bipolar orientations}.
Let $\G$ be a finite subgraph of $\Z^2$, that is, the part of $\Z^2$ bounded by a rectilinear integer polygon.
Let $N$ and $S$ be distinct vertices on the outer boundary of $\G$. A \emph{bipolar orientation}
is an orientation of the edges of $\G$ which is acyclic (has no oriented cycles), has only one source, at $N$, and has only
one sink, at $S$.

We give a bijection between bipolar orientations of $\G$ and
6-vertex configurations on another graph $H$, the ``double'' of $\G$,
whose vertices are the vertices and faces of $\G$,
with edges of $H$ connecting vertices of $\G$ to their incident faces of $\G$.
Edges of $\G$ correspond to faces of $H$.  (See Fig.~\ref{bpfig}a.)

At each vertex $v$ of $\G$, the outgoing edges in the bipolar orientation form a contiguous interval in the circular order
around $v$, that is, there are no vertices for which the orientation is in-out-in-out. Equivalently
the incoming arrows form a contiguous interval around $v$.
In the corresponding 6-vertex configuration, outgoing arrows from $v$ point
to the two faces that separate these intervals.  For each face~$f$ of $\G$,
the bipolar orientation restricted to that face has a unique source and unique sink; the 6-vertex
arrows point from this face to these two extremal vertices.  It is easy to check that each
edge of $H$ is oriented by precisely one of these two rules,
so it has out-degree 2 everywhere, that is, it is a 6-vertex configuration.

Given an edge in a bipolar-oriented graph $\G$, there is a canonical path to the sink, obtained by travelling along that edge
in the direction of its orientation and, when arriving at a vertex, taking the maximally left outgoing edge from the new vertex.
The union of these paths forms a tree, the ``SE-tree'', drawn in blue in Fig.~\ref{bpfig}b.
The analogous ``NW-tree'', which is the SE-tree for the bipolar orientation
obtained by reversing all the arrows, is drawn in red in Fig.~\ref{bpfig}c.
The SE-tree and NW-tree do not cross each other, so there is a curve winding between them, which is
shown in green in Fig.~\ref{bpfig}d.
This map from bipolar orientations to Peano curves was first described for general planar graphs in \cite{KMSW1}.
This Peano curve is the same curve defined by the 6-vertex height function.

\newlength\figwidth
\newcommand\underfig[2]{\settowidth{\figwidth}{#1}\hspace{0.5\figwidth}\clap{#1}\clap{#2}\hspace{0.5\figwidth}}
\newcommand{\ymin}{0.2}
\newcommand{\ymax}{1.8}
\begin{figure*}
\mbox{%
\hspace{-6pt}%
\underfig{\raisebox{10pt}{\begin{tikzpicture}[x=1.2cm,y=1.9cm]
\draw (0,\ymin) -- (4,\ymin);
\draw (0,\ymin) -- (0,\ymax);
\foreach\x in {0,0.2,...,4} \draw (\x,\ymin+0.015)--(\x,\ymin-0.015);
\foreach\x in {0,...,4} {\draw (\x,\ymin+0.03)--(\x,\ymin-0.03); \draw (\x,\ymin) node [below] {\scalebox{0.8}{$\x$}};}
\foreach\y in {0.2,0.3,...,1.8} \draw (0,\y) node {\rule{2pt}{.5pt}};
\foreach\y in {0.5,1.0,1.5} {\draw (0,\y) node {\rule{4pt}{.5pt}}; \draw (0,\y) node [left] {\scalebox{0.8}{$\y$}};}
\draw (4,\ymin) node [right] {\scalebox{0.9}{$C^2$}};
\draw (0,\ymax) node [above] {\scalebox{0.9}{winding}};
\clip (0,\ymin) rectangle (4,\ymax);
\draw plot[only marks, mark=*, mark options={blue, mark size=.5}] file {6v-path-winding-mc.data};
\draw [color=red, thick, domain=0.01:4,samples=400] plot (\x, {45/acos(1-\x/2)});
\end{tikzpicture}}}{(a) winding angle variance}
\renewcommand{\ymin}{0.16}
\renewcommand{\ymax}{0.65}
\underfig{\raisebox{10pt}{\begin{tikzpicture}[x=1.2cm,y=6.20408cm]
\draw (0,\ymin) -- (4,\ymin);
\draw (0,\ymin) -- (0,\ymax);
\foreach\x in {0,0.2,...,4} \draw (\x,\ymin) node {\rule{.5pt}{2pt}};
\foreach\x in {0,...,4} {\draw (\x,\ymin) node {\rule{.5pt}{4pt}}; \draw (\x,\ymin) node [below] {\scalebox{0.8}{$\x$}};}
\foreach\y in {0.2,0.25,...,0.6} \draw (0,\y) node {\rule{2pt}{.5pt}};
\foreach\y in {0.2,0.4,0.6} {\draw (0,\y) node {\rule{4pt}{.5pt}}; \draw (0,\y) node [left] {\scalebox{0.8}{$\y$}};}
\draw (4,\ymin) node [right] {\scalebox{0.9}{$C^2$}};
\draw (0,\ymax) node [above] {\scalebox{0.9}{winding}};
\clip (0,\ymin) rectangle (4,\ymax);
\draw plot[only marks, mark=*, mark options={blue, mark size=.5}] file {6v-path-outer-winding-mc.data};
\draw [color=red, thick, domain=0.01:4,samples=400] plot (\x, {1/(1+acos(1-\x/2)/90*(1+sqrt(1+180/acos(1-\x/2))))});
\end{tikzpicture}}}{(b) outer boundary winding angle variance}
\renewcommand{\ymin}{1.07}
\renewcommand{\ymax}{1.32}
\underfig{\raisebox{10pt}{\begin{tikzpicture}[x=1.2cm,y=12.16cm]
\draw (0,\ymin) -- (4,\ymin);
\draw (0,\ymin) -- (0,\ymax);
\foreach\x in {0,0.2,...,4} \draw (\x,\ymin) node {\rule{.5pt}{2pt}};
\foreach\x in {0,...,4} {\draw (\x,\ymin) node {\rule{.5pt}{4pt}}; \draw (\x,\ymin) node [below] {\scalebox{0.8}{$\x$}};}
\foreach\y in {1.08,1.1,...,1.32} \draw (0,\y) node {\rule{2pt}{.5pt}};
\foreach\y in {1.1,1.2,1.3} {\draw (0,\y) node {\rule{4pt}{.5pt}}; \draw (0,\y) node [left] {\scalebox{0.8}{$\y$}};}
\draw (4,\ymin) node [right] {\scalebox{0.9}{$C^2$}};
\draw (0,\ymax) node [above] {\scalebox{0.9}{$D_f$}};
\clip (0,\ymin) rectangle (4,\ymax);
\draw plot[only marks, mark=*, mark options={blue, mark size=.5}] file {6v-path-outer-dimension-mc.data};
\draw [color=red, thick, domain=0.01:4,samples=400] plot (\x, {1+0.5/(1+acos(1-\x/2)/90*(1+sqrt(1+180/acos(1-\x/2))))});
\end{tikzpicture}}}{(c) outer boundary dimension}\hspace{24pt}}
\caption{Monte Carlo estimates (points, using $L=256$ and $L=512$) and SLE predictions (curves, using \eqref{eqn:kappa'}, \eqref{eqn:winding}, \eqref{eqn:outer-winding}, and \eqref{eqn:outer-dimension}) of (a) the winding angle variance coefficient, (b) outer boundary winding angle variance coefficient, and (c) outer boundary dimension of the 6-vertex model's Peano curve, as a function of $C^2$.  The apparent deviation in the right figure represents a finite size effect that we believe would go away on larger grids. Each dot represents an independent winding angle variance or dimension estimate for a certain $C^2$ value.  Since these quantities are continuous in $C^2$, and independent estimates
for many nearby values of $C^2$ are given, the vertical fluctuations in the points effectively serve as error bars.
}
\label{fig:monte-carlo-wind-dim}
\end{figure*}
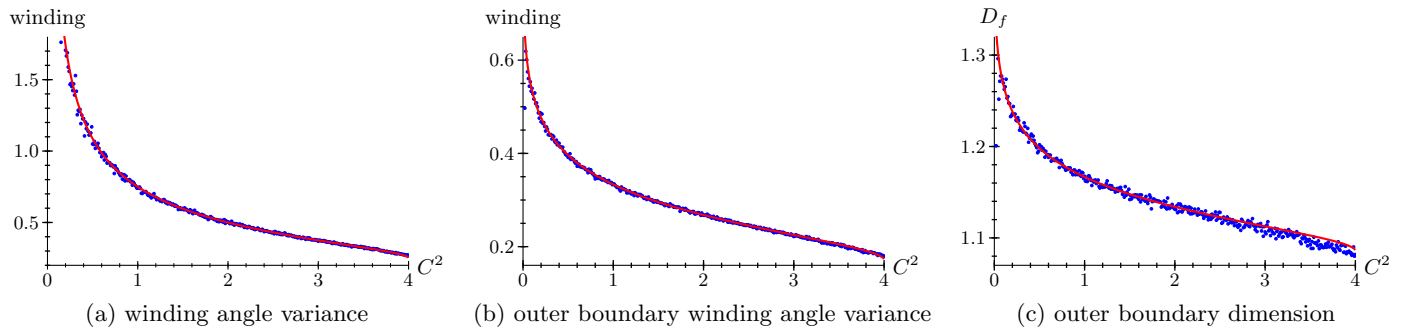

Fig.~\ref{fig:ice-tour} shows a random sample of the Peano curve associated to a large square ice configuration on the square grid.
For planar graphs, perfect samples for the 6-vertex models with $C\ge 1$
can be obtained from single-site Glauber dynamics and coupling from the past
\cite{propp-wilson:cftp}.

\medskip
\bookmark[dest=monte-carlo]{Monte Carlo simulations}
\hypertarget{monte-carlo}{\noindent\textit{\textbf{Monte Carlo simulations.}}}
We used Monte Carlo simulations to check that the 6-vertex model Peano curve is described by $\SLE_{\kappa'}$.  We produced 6-vertex configurations on an $L\times L$ torus for various values of $L$, to eliminate boundary effects.  We measured the winding angle variance of the Peano curve, and also the dimension of the outer boundary of the Peano curve.

SLE theory predicts that the Peano curve's winding angle variance scales as
\vspace{-0.5pt}
\begin{equation} \label{eqn:winding}
\frac{4\kappa'}{(\kappa'-4)^2}\, \ln L\,.
\end{equation}
\vspace{-0.5pt}\cite{miller-sheffield:ig4}.  Since the winding of the curve is
given by the height function,
we measured the height function variance.

The outer boundary corresponds to paths within the blue SE-tree in Fig.~\ref{bpfig}b.  Since the simulations are done on a torus, the ``SE-tree'' is actually a cycle-rooted spanning forest (CRSF), and we measured both the winding angle variance and the length~$\ell$ of the cycle in the cycle-rooted spanning tree containing the edge at the origin.  The SLE prediction is that the outer boundary's winding angle variance scales as
\begin{equation} \label{eqn:outer-winding}
\frac{4}{\kappa'}\, \ln L\,,
\end{equation}
and that its
length scales as $\ell\sim L^{D_{\!f}}$ where
\begin{equation} \label{eqn:outer-dimension}
D_{\!f} = 1+\kappa/8 = 1+2/\kappa'\,.
\end{equation}
We estimated the winding angle variance coefficients and the outer boundary dimension using samples for $L=256$ and $L=512$, as shown in Fig.~\ref{fig:monte-carlo-wind-dim}.

The estimates for the winding angle variance coefficient is an excellent fit to the predicted value.  Since the formula relating $\kappa'$ to $C^2$ was
derived from Nienhuis' formula \eqref{eqn:g-C}, the left panel of Fig.~\ref{fig:monte-carlo-wind-dim} is essentially an experimental verification of Nienhuis' formula.

The outer boundary winding angle variance and dimension estimates (middle and right panels of Fig.~\ref{fig:monte-carlo-wind-dim}) are both independent tests of the curve's convergence to SLE.  The estimated values are a close match to the predicted value, though when $C^2\approx 3.5$, the measured dimension deviates from the prediction by as much as $0.015$.  Further tests of the distribution of the loop length~$\ell$ and its dependence on $L$
suggest that the convergence to the asymptotic behavior occurs for larger values of $L$ when $C^2\approx 4$ than when, for example, $C^2\approx 2$.  Overall, the experiments are consistent with convergence to $\SLE$.

\medskip
\bookmark[dest=acknowledgments]{Acknowledgments}
\hypertarget{acknowledgments}{\noindent\textit{\textbf{Acknowledgments.}}}
R.K.\ was supported by an NSF grant and a Simons Foundation grant.
J.M.\ was supported by an NSF grant.
S.S.\ was supported by a Simons Foundation grant,
an NSF grant, and two EPSRC grants.

\def\@rst #1 #2other{#1}
\newcommand\MR[1]{\relax\ifhmode\unskip\spacefactor3000 \space\fi
  \MRhref{\expandafter\@rst #1 other}{#1}}
\newcommand{\MRhref}[2]{\href{http://www.ams.org/mathscinet-getitem?mr=#1}{MR#1}}

\newcommand{\arXiv}[1]{\href{http://arxiv.org/abs/#1}{arXiv:#1}}
\newcommand{\arxiv}[1]{\href{http://arxiv.org/abs/#1}{#1}}
\renewcommand \doibase [0]{http://doi.org/}

\bookmarksetup{startatroot}
\bibliography{../bipolar,../ice,../../exploration/activity/activity}

\end{document}